\documentclass[aps,twocolumn,showpacs,superscriptaddress,amsmath,amssymb,amsfonts,floatfix,longbibliography]{revtex4-1}
\usepackage[T1]{fontenc} 
\usepackage{float}
\usepackage{dcolumn}
\usepackage{bm}
\usepackage{amssymb}
\usepackage{microtype}
\usepackage{xfrac}
\usepackage{gensymb}
\usepackage{multirow}
\usepackage{enumitem}
\usepackage{natbib,hyperref}
\usepackage{graphicx}
\usepackage[normalem]{ulem}

\begin{document}

\title{Short-range order and local distortions in entropy stabilized oxides}

\author{Solveig S. Aamlid}
    \affiliation{Stewart Blusson Quantum Matter Institute, University of British Columbia, Vancouver, BC V6T 1Z4, Canada}

\author{Sam Mugiraneza}
    \affiliation{Stewart Blusson Quantum Matter Institute, University of British Columbia, Vancouver, BC V6T 1Z4, Canada}
    \affiliation{Department of Chemistry, University of British Columbia, Vancouver, BC V6T 1Z1, Canada}

\author{Mario U. Gonz\'alez-Rivas}
    \affiliation{Stewart Blusson Quantum Matter Institute, University of British Columbia, Vancouver, BC V6T 1Z4, Canada}
    \affiliation{Department of Physics \& Astronomy, University of British Columbia, Vancouver, BC V6T 1Z1, Canada}

\author{Graham King}
    \affiliation{Canadian Light Source, Saskatoon, Saskatchewan S7N 2V3, Canada}

\author{Alannah M. Hallas}
\email[Email: ]{alannah.hallas@ubc.ca}
    \affiliation{Stewart Blusson Quantum Matter Institute, University of British Columbia, Vancouver, BC V6T 1Z4, Canada}
    \affiliation{Department of Physics \& Astronomy, University of British Columbia, Vancouver, BC V6T 1Z1, Canada}
    \affiliation{Canadian Institute for Advanced Research (CIFAR), Toronto, ON, M5G 1M1, Canada}
    
\author{J\"org Rottler}
\email[Email: ]{jrottler@physics.ubc.ca}
    \affiliation{Stewart Blusson Quantum Matter Institute, University of British Columbia, Vancouver, BC V6T 1Z4, Canada}
    \affiliation{Department of Physics \& Astronomy, University of British Columbia, Vancouver, BC V6T 1Z1, Canada}

\date{\today}

\begin{abstract}
\smallskip
\begin{center}
\normalsize\textbf{ABSTRACT}\\    
\end{center}
An idealized high entropy oxide is characterized by perfect chemical disorder and perfect positional order. In this work, we investigate the extent to which short-range order (SRO) and local structural distortions impede that idealized scenario. 
Working in the entropy stabilized $\alpha$-PbO$_2$ structure, we compare a two-component system, (Ti,Zr)O$_2$, with a four-component system, (Ti,Zr,Hf,Sn)O$_2$, using a combination of experimental and computational approaches. 
Special quasi-random structures are used in conjunction with density functional theory calculations to investigate the local distortions around specific elements revealing significant local distortions that are relatively insensitive to the number of chemical constituents. 
Using finite temperature Monte Carlo simulations, we are able to reproduce the previously experimentally observed SRO and transition temperature for the two-component system. However, the ideal configurational entropy is never reached, so SRO is expected even at synthesis temperatures. On the other hand, the order-disorder transition temperature is dramatically lower and experimentally inaccessible for the four-component system, while the configurational entropy is closer to ideal and less sensitive to temperature.
Total scattering measurements and pair distribution function analysis of slow-cooled and quenched samples support this view. In general, we demonstrate that SRO effects in high entropy materials are less prevalent as more components are added in, provided the pairwise interaction strengths remain comparable, while local distortions are less affected by the number of components.
\end{abstract}

\maketitle

\section{Introduction\label{sec:Introduction}}
 
High entropy materials are crystalline compounds in which different elements share one or more crystallographic lattice sites~\cite{george2019high,oses2020high}. Initially, it may be assumed that all the chemical elements in such a material are populating the lattice in a perfectly random manner while the lattice itself remains perfectly undistorted. However, in real materials, this assumption quickly breaks down due to the formation of chemical short-range order (SRO) and local distortions to the crystal lattice, both of which can lower the free energy~\cite{Aamlid2023Understanding}.  Recently, there has been a surge of research on characterizing and quantifying these effects in high entropy materials \cite{Ferrari2023}, particularly in cases where they can be used as a control parameter for functional properties \cite{Lun2021, Bajpai2022, Berardan2017}.

SRO and local distortions are notoriously difficult to detect due to the short length scales and complexity of possible configurations involved, and that difficulty increases with the number of chemical elements or lower crystallographic symmetry~\cite{keen2015crystallography}. 
Prime examples of the presence and influence of SRO come from the high entropy alloy community, where the effect of SRO on microstructural and mechanical properties is of particular interest~\cite{Wu2021}. Work in the alloy field also illustrates that heat treatments can influence the magnitude of SRO, which in turn may control the functional properties, exemplified in regards to the magnetic properties of CoCrNi~\cite{Bajpai2022}.  
Local distortions, meanwhile, are of greater interest in oxide materials (or other compounds with counterion sublattices), since the oxygen sublattice is typically more malleable and acts as a mediator for size mismatch.
The magnitude of local distortions may, much like SRO, also be affected by thermal treatments and govern the functional properties, as is the case of the Jahn-Teller distortion around Cu in the prototypical rock-salt high entropy oxide (Mg,Co,Ni,Cu,Zn)O~\cite{Berardan2017}. The notation where multiple elements are separated with commas in parenthesis is used throughout this work to indicate equimolarity and solid solution on a single lattice site.

\begin{figure*}[htb]
  \centering
  \includegraphics[width=0.9\textwidth]{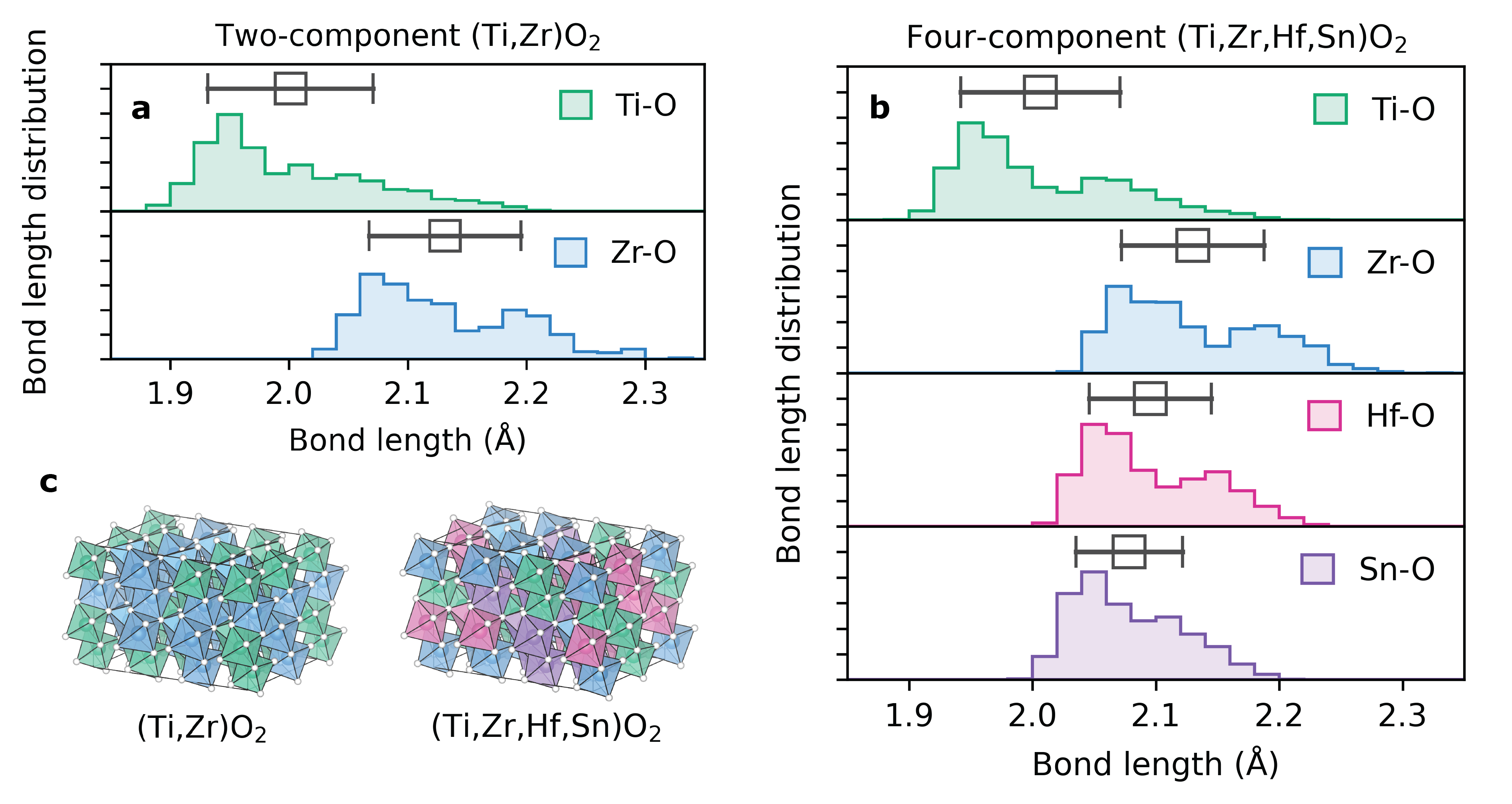}
  \caption{\textbf{Cation-oxygen bond lengths calculated with density functional theory.} (a) The distributions of Ti-O and Zr-O bond lengths in the first coordination sphere of (Ti,Zr)O$_2$ obtained by averaging over 3 special quasi-random structure (SQS) cells. The average and standard deviation for each distribution are shown in black. (b) The distributions of Ti-O, Zr-O, Sn-O, and Hf-O bond lengths in the first coordination sphere of (Ti,Zr,Hf,Sn)O$_2$ obtained by averaging over 28 SQS cells. The average and standard deviation for each distribution are shown in black. (c) Representative relaxed SQS cells of the two- and four-component systems reveal that the octahedral coordination and connectivity are preserved even in $P1$ symmetry.}
  \label{fgr:SQS}
\end{figure*}

In this work, we use a combination of experimental and computational techniques to study the evolution of SRO and local distortions upon increasing the chemical complexity from a two-component system to a four-component system, thereby increasing the configurational entropy while also modifying the enthalpic landscape.
The four-component system is (Ti,Zr,Hf,Sn)O$_2$, an entropy-stabilized oxide that crystallizes in the orthorhombic $\alpha$-PbO$_2$ crystal structure~\cite{Aamlid2023Phase,He2021fourcomponent}. The two-component model system is (Ti,Zr)O$_2$, which crystallizes in the same crystal structure and is likewise entropy stabilized~\cite{Hom2001}. While Ti and Zr are the two elements from the four-component system that are most different in ionic radii, they are closer in terms of mass, electronic band gap in their binary oxide, polarizability, and electronegativity than some of the other possible cation pairings. (Ti,Zr)O$_2$ 
displays well-characterized local distortions and SRO which are known to be influenced by thermal treatments \cite{Mchale1986, Christoffersen1992, Wang1997, troitzsch2005crystal}. It is also interesting to note that this system seems to prefer SRO over a reversible phase transition back to its constituent oxides~\cite{Mchale1986}, meaning that the SRO may be the byproduct of inhibited long-range order. 

DFT calculations are used to assess local distortions in (Ti,Zr)O$_2$ and (Ti,Zr,Hf,Sn)O$_2$ in the perfectly chemically disordered limit. While varying cation-oxygen distances and cation off-centering are observed, the magnitudes of these are similar between the two- and four-component compounds. A truncated cluster expansion is then performed to find effective cluster interaction energies for every nearest neighbor cation pair, and those energies are subsequently used as inputs in Metropolis Monte Carlo simulated annealing runs. These results allow for direct access to the nature of the SRO, the non-ideal configurational entropy, and the order-disorder phase transition temperature. Finally, x-ray total scattering measurements and pair distribution function analysis validate the computational findings, confirming that SRO is inhibited by increasing chemical complexity, while local distortions are less affected.

\section{Results and discussion}

\subsection{Local distortions from DFT calculations of SQS cells}

In order to investigate their local distortions, 144 atom special quasi-random structure (SQS) cells of (Ti,Zr)O$_2$ (3 initial configurations) and (Ti,Zr,Hf,Sn)O$_2$ (28 initial configurations) were relaxed using DFT. Both of these materials are observed experimentally to crystallize in the $\alpha$-PbO$_2$ structure, which is an orthorhombic structure (space group $Pbcn$) made up of edge- and corner-sharing octahedra. In the average structure, the octahedral oxygen environment is distorted to have three inequivalent oxygen-metal bond distances that differ by up to 0.25~\AA. Representative relaxed SQS cells for each of these materials are shown in Figure \ref{fgr:SQS}(c), indicating that the general connectivity of the $\alpha$-PbO$_2$ structure is preserved regardless of the global $P1$ symmetry. Large cation-dependent local distortions are observed and histograms showing the frequency of different bond lengths for the shortest cation-oxygen distance in the two different compounds are displayed in Figure \ref{fgr:SQS}(a) for (Ti,Zr)O$_2$ and (b) for (Ti,Zr,Hf,Sn)O$_2$. The mean and standard deviation is shown for each cation-oxygen pair, and reproduced in Table \ref{tab:SQS}. Calculations where the SQS cell was initialized with some degree of short range order were attempted, and these local distortions are not strongly influenced by intermediate degrees of SRO.

In the case of (Ti,Zr)O$_2$, the mean values for the Ti-O and Zr-O bond distances are 2.00(7)~\AA \, and 2.13(6)~\AA \, respectively, and an overall broader distribution is obtained for Ti. The smaller average bond distance for Ti reflects its smaller Shannon radii, which causes the local oxygen environment to contract accordingly. The experimentally determined cation-oxygen distance in (Ti,Zr)O$_2$ is 2.04~\AA\, at room temperature from x-ray diffraction~\cite{troitzsch2005crystal}, which is in line with an average of the two environments. However, standard x-ray diffraction is insensitive to distortions of the local environment, as without elemental sensitivity, only the average bond distances can be refined. We can instead compare these values with those determined via extended x-ray absorption fine structure (EXAFS) measurements, a technique that has elemental sensitivity, which was performed by Xu et al. \cite{Xu2000solgel}. The bond distances determined by EXAFS are shorter than the ones obtained from XRD or DFT, but the difference between the Ti-O and Zr-O bonds is consistent. The tetravalent and six-coordinated Ti-O and Zr-O bond distances from Shannon are 2.005 and 2.12~\AA\ respectively, again in line with the results from DFT \cite{Shannon1976}. DFT calculations are hence able to capture the local distortions in the two-component system and can be applied to the four-component system.

The bond distance histograms for the four-component sample reveal a similar tendency towards cation-dependent distortions of the local oxygen environment. Specifically, the Ti-O and Zr-O bond distance distributions relax to practically identical distributions as the ones obtained for the two-component compound. The average Hf-O and Sn-O bond distances are closer to the overall average bond distance, resulting in smaller standard deviations for these cations.

\begin{table}[tbp]
\caption{Comparison of experimental and computational average cation-oxygen bond length in the first coordination shell for the two- and four-component compounds. The experimental values come from EXAFS and are reproduced from Xu et al. \cite{Xu2000solgel}, with the standard deviation representing the convolution of thermal motion, positional disorder, and different bond directions. The computational values come from the distributions calculated by DFT shown in Figure \ref{fgr:SQS}(a,b). The bracketed number represents the standard deviation in the last digit.}
\noindent
\begin{tabular*}{\columnwidth}{@{\extracolsep{\stretch{1}}}*{4}{c}@{}}
\toprule
&\multicolumn{1}{c}{(Ti,Zr)O$_2$}&\multicolumn{1}{c}{(Ti,Zr)O$_2$}&\multicolumn{1}{c}{(Ti,Zr,Hf,Sn)O$_2$}\\
&\multicolumn{1}{c}{EXAFS}&\multicolumn{1}{c}{DFT}&\multicolumn{1}{c}{DFT}\\
\hline
Ti-O                   & 1.94(-)~\AA    & 2.00(7)~\AA    & 2.00(6)~\AA  \\
Zr-O                   & 2.07(8)~\AA     & 2.13(7)~\AA  & 2.13(6)~\AA \\
Hf-O                                   &  &           & 2.08(4)~\AA \\
Sn-O                               &  &           & 2.10(5)~\AA   \\
\toprule
\end{tabular*}
\label{tab:SQS}
\end{table}

A marked asymmetry of the cation-oxygen bond lengths can be found across all cations in both compounds, where the distribution is peaked at lower distances with a long tail at higher distances. The origin of this asymmetry is two-fold. The octahedra that comprise the $\alpha$-PbO$_2$ structure are partly edge-sharing and partly corner-sharing, and cations tend to move away from shared edges. In the \textit{Pbcn} symmetry, there are three inequivalent cation-oxygen bond lengths, and the equidistant bonds are on the same edge of the octahedron, implying a certain off-centering in the octahedra also in a globally ordered crystal. The polar moment that may arise from this is perfectly canceled by nearby ions. This structure is hence antipolar and is one of the possible parent phases for the ferroelectric HfO$_2$ thin films \cite{Raeliarijaona2023}. The off-centering from this effect is expected to be cooperative, since the edge-sharing is equally present in the heavily disordered compound. 

In addition to the asymmetry from the edge-sharing, there is also a propensity for some of these cations to off-center due to second-order Jahn-Teller (SOJT) effects. Of the ions present in this study, Ti$^{4+}$ has the largest propensity for the SOJT, Zr$^{4+}$ and Hf$^{4+}$ have a weaker propensity, while Sn$^{4+}$ is not Jahn-Teller active. This is reflected in the degree of asymmetry for the different ions. The EXAFS data for the two-component sample from literature corroborate the view of the particularly off-centered Ti ion, as the coordination number that fits the data best is less than six \cite{Xu2000solgel}. Any additional off-centering of the cation inside an octahedron can lead to a small local polar moment from a second-order Jahn-Teller distortion since all the ions are $d^0$ and a large band gap is expected. This could open up the possibility for a polar glass behavior, although the \textit{Pbcn} space group remains a macroscopically non-polar structure.

\begin{figure*}[htbp]
  \centering
  \includegraphics[width=1\textwidth]{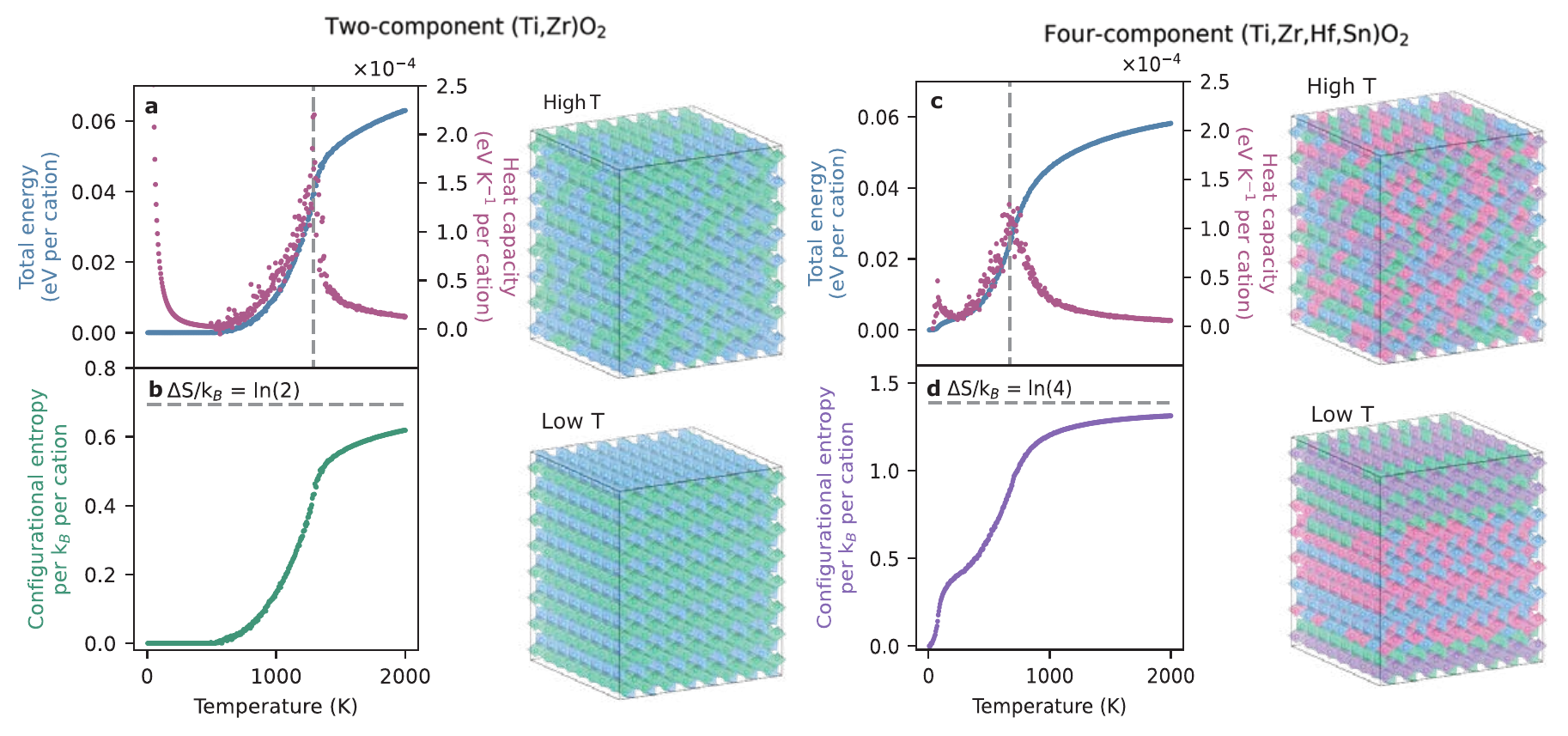}
  \caption{\textbf{Short-range ordering from simulated annealing.} The results shown in the left-hand set of panels are for (Ti,Zr)O$_2$ while the right-hand set of panels are for (Ti,Zr,Hf,Sn)O$_2$. (a,c) The total energy (blue, left axes) and heat capacity (pink, right axes) as a function of temperature reveal a thermodynamic transition arising from cation order at 1290~K and 670~K for the two- and four-component materials, respectively. The resulting cation ordering can be visualized by inspecting the high (2000~K) and low (5~K) temperature Monte Carlo cells where the cations polyhedra for Ti (green), Zr (blue), Hf (pink), and Sn (purple) are shown and oxygen omitted for clarity. (b,d) Temperature dependence of the configurational entropy determined by thermodynamic integration of the calculated heat capacity. The configurational entropy falls short of the perfectly random limit represented by the grey dashed line even up to 2000~K, especially for the two-component sample.}
  \label{fgr:MC}
\end{figure*}

Beyond the bond distances and cation off-centering, we can also quantify the distribution of octahedral distortions by considering the continuous symmetry measure (CSM) value for the $M$O$_6$ octahedra in our SQS cells. CSM is a measure of similarity between shapes, where a perfect polyhedron gets a score of zero while increasing distortion leads to progressively larger CSM scores~\cite{pinsky1998continuous}. As a reference, we can take pure TiO$_2$ and ZrO$_2$ in the $\alpha$-PbO$_2$ structure, which have CSM values of 0.46 and 1.04 respectively, illustrating that the octahedron in this structure is inherently distorted. The chemically resolved results for the 144 atom SQS cells for the two-component and four-component compounds can be found in Table \ref{tab:CSM}. The degree of octahedral distortion of the TiO$_6$ and ZrO$_6$ environments is slightly larger in the SQS cells than in the binary oxides while the distribution of distortions, which can be inferred from the standard deviations, is smaller in the four-component (Ti,Zr,Hf,Sn)O$_2$ than in the two-component (Ti,Zr)O$_2$. Octahedra with Zr or Hf are more distorted than octahedra with Ti or Sn, a result that appears consistent with the respective ground state crystal structures of the cations. The baddeleyite structure of ZrO$_2$ and HfO$_2$ has the cations in a highly distorted seven-coordinated environment while the rutile structure of TiO$_2$ and SnO$_2$ is built up of slightly elongated octahedra. This result suggests that, similar to bond length, the average degree of distortion is characteristic of the cation in question but that adding additional chemical components enables a more uniform distortion.

\begin{table}[tbp]
\caption{Chemically resolved continuous symmetry measure (CSM) values for the degree of octahedral distortion in the two-component (Ti,Zr)O$_2$ and four-component (Ti,Zr,Hf,Sn)O$_2$ SQS cells as compared to pure TiO$_2$ and ZrO$_2$ in the $\alpha$-PbO$_2$ structure. The bracketed number represents the standard deviation in the last digit.}
\noindent
\begin{tabular*}{\columnwidth}{@{\extracolsep{\stretch{1}}}*{4}{c}{c}{c}@{}}
\toprule
&TiO$_2$& ZrO$_2$& (Ti,Zr)O$_2$ & (Ti,Zr,Hf,Sn)O$_2$ \\
\hline
TiO$_6$ distortion &0.46&-& 0.8(3) & 0.7(1) \\
ZrO$_6$ distortion &-&1.04& 1.4(4) & 1.4(2)  \\
HfO$_6$ distortion &-&-& - & 1.1(2) \\
SnO$_6$ distortion &-&-& - & 0.8(2)  \\ 
\toprule
\end{tabular*}
\label{tab:CSM}
\end{table}

\subsection{SRO from cluster expansion and Monte Carlo simulations}

A truncated version of the cluster expansion framework \cite{SANCHEZ1984334} was used to study the SRO in the two- and four-component compounds from a computational standpoint. Since the type of ordering and ordering temperature is well known for (Ti,Zr)O$_2$, it is used as a test system to verify the validity of this approach. Tables of the effective cluster interactions can be found in the supporting information. 
Simulated annealing based on a Metropolis Monte Carlo algorithm was performed based on ECI parameters from the truncated cluster expansion. The resulting total energy and heat capacity are shown in Figure \ref{fgr:MC}(a) for the two-component system. The broad order-disorder phase transition defined by the peak in the heat capacity at 1290~K for (Ti,Zr)O$_2$ is in  satisfactory agreement with experimental data where the ordering is reported between 1333 and 1433 K. The qualitative layering of Ti and Zr along the $a$-axis is also in agreement with available experimental data \cite{Mchale1986, Christoffersen1992, troitzsch2005crystal}.

Thermodynamic integration (see Methods) was used to find the non-ideal configurational entropy, which is difficult to access experimentally. The ideal configurational entropy for a two-component system is $k_B \ln(2)$ per cation~\cite{Aamlid2023Understanding}, representing complete random disorder of Ti and Zr across the shared cation site, serves as an upper bound. In Figure~\ref{fgr:MC}(b), the configurational entropy is shown for the two-component compound. Interestingly, even at the synthesis temperature of 1673 K, the configurational entropy is not saturated, indicating that, under experimentally relevant conditions, (Ti,Zr)O$_2$ is never actually fully disordered. Representative cation configurations at high (2000~K) and low (5~K) temperatures are also presented in Figure~\ref{fgr:MC}. A conventional cluster expansion for the two-component system can be found in the supporting information. This cluster expansion has similar layering along the $a$ direction, but the Ti-layers are doubled and there is a dissolution of ZrO$_2$. This double Ti layers are also observed in the mineral Srilankite (Ti$_2$ZrO$_6$)~\cite{troitzsch2005crystal}.  

The same procedures were followed for the four-component composition, and the corresponding results are shown in Figure \ref{fgr:MC}(c) and (d). The cation ordering phase transition temperature is moved downward in temperature to 670 K. The ECIs are of similar magnitudes to the two-component system, indicating that the mixing enthalpy and other energetic effects are similar in the two systems. However, the ideal configurational entropy is now $k_B \ln(4)$ per cation twice as much as before, which causes the phase transition temperature to go down to roughly half of what it was. The thermodynamic integration in Fig.~\ref{fgr:MC}(d) shows that the ideal configurational entropy is not achieved up to 2000~K, although it is closer than it was for the two-component system and the slope is smaller. It should be noted that the apparent phase separation and ordering which appear below 670 K in the calculations cannot realistically be observed experimentally due to kinetic constraints. Metropolis Monte Carlo has no kinetic information and finds the thermodynamic ground state, while the experimental cation diffusion at 670~K is very slow.

\begin{figure}[htbp]
  \centering
  \includegraphics[width=0.9\columnwidth]{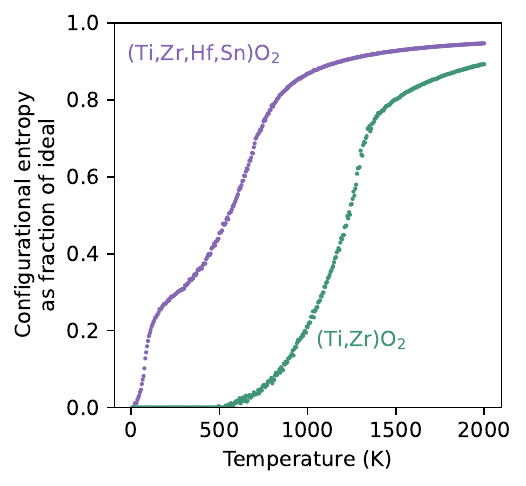}
  \caption{\textbf{Configurational entropy as a fraction of ideal entropy.} The configurational entropy calculated from the Monte Carlo simulations using thermodynamic integration is displayed as a fraction of the ideal entropy ($k_B\ln(2)$ per cation for (Ti,Zr)O$_2$ and $k_B\ln(4)$ per cation for (Ti,Zr,Hf,Sn)O$_2$) as a function of temperature. }
  \label{fgr:smallconfigentropy}
\end{figure}

In order to appreciate the effect of changing the number of constituents, we plot the configurational entropy normalized with respect to the ideal entropy for both the two and four-component materials in Figure~\ref{fgr:smallconfigentropy}. The entropy release associated with the cation ordering transition shifts significantly downward in temperature in the four-component material compared to the two-component material. At the synthesis temperature of 1673 K, the four-component configurational entropy has reached 94\% of the ideal value while the two-component has reached 85\%. We therefore expect a higher degree of short-range cation ordering to be present in the two-component system compared to the four-component system at all realistic synthesis conditions.

\begin{figure*}[tbp]
  \centering
  \includegraphics[width=\textwidth]{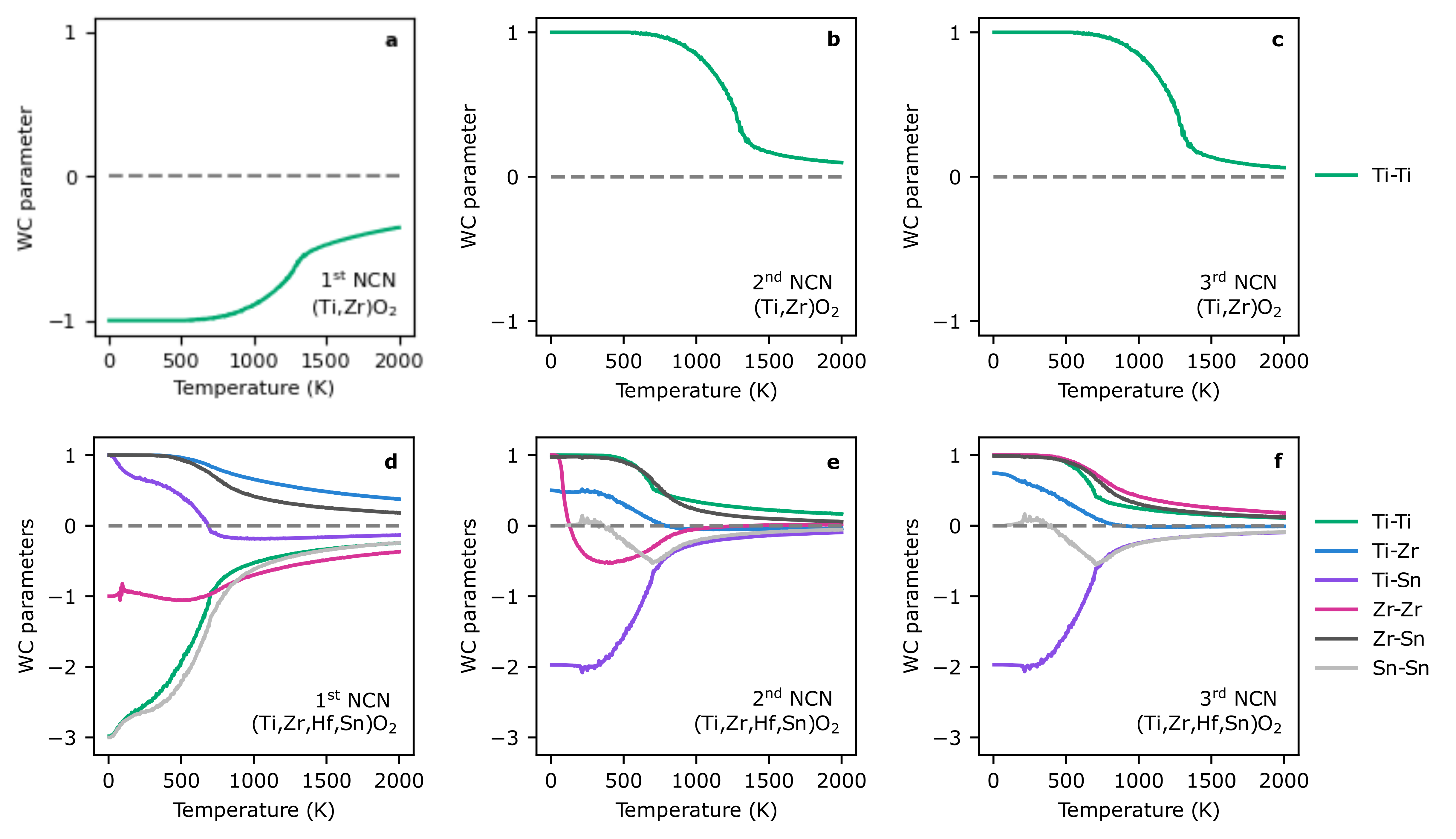}
  \caption{\textbf{Warren-Cowley (WC) parameters from simulated annealing} in (a), (b) and (c) for the Ti-Ti interaction for the 1$^{\mathrm{st}}$, 2$^{\mathrm{nd}}$, and 3$^{\mathrm{rd}}$ nearest cation neighbor (NCN) clusters respectively in the two-component, and in (d), (e) and (f) for the six unique cation pairings Ti-Ti, Ti-Zr, Ti-Sn, Zr-Zr, Zr-Sn, and Sn-Sn) for the NCNs in the four-component system. Significant deviations from perfectly random disorder, which would be indicated by a WC parameter of 0, are observed for the two-component system, even at 2000 K while smaller deviations are found for the four-component.  }
  \label{fgr:WC}
\end{figure*}

In order to quantify the degree of SRO, we employ the commonly used Warren-Cowley (WC) parameters~\cite{cowley1950approximate,cowley1965short}
\begin{equation}
\alpha_{ij}=1-\frac{P_{ij}}{2c_ic_j},
\end{equation}
where $P_{ij}$ denotes the probability of observing a pair of species $i$ and $j$ in a given configuration, and $c_i$ refers to the average concentration of species $i$. In the disordered phase, $\alpha_{ij}=0$. In Fig.~\ref{fgr:WC}, we report these parameters for the three nearest cation neighbor (NCN) clusters. For two component (Ti,Zr)O$_2$ (panels (a-c)), the WC parameters. plotted for the Ti-Ti pair, take values between $-1$ and $1$, where $\alpha_{ij}=-1$ means that the two indexed cations are fully ordered along that direction while $\alpha_{ij}=1$ denotes the absence of that particular pairing in a given direction. For (Ti,Zr)O$_2$ there is residual order in the first NCN even at 2000~K, while perfect layered order is observed below 700~K corresponding to alternating layers of Ti and Zr along the $a$-axis.

In four-component (Ti,Zr,Hf,Sn)O$_2$  (panels (d-f)), the differing number of species and their relative concentrations modifies the range of the WC parameters, plotted for 6 independent pairs, span $-3$ to $1$ for full long range order and $0$ indicating a fully random configuration. At 2000~K, the level of SRO is reduced as compared to the two-component material but the WC parameters have not fully reached zero in all clusters, corroborating the findings from the integrated entropy. Ti-Ti interactions behave similarly in all NCNs and are predicted to order strongly below 700~K. Other pairs, such as Sn-Sn, exhibit a weaker ordering tendency with some remaining highly disordered (close to $\alpha_{ij}=0$) in the second and third NCN in the zero temperature limit. In both cases, Ti forms layers along the same direction, but the smaller WC parameter at high temperatures show that the short-range order is suppressed in the four-component material for Ti-Ti as well. Moreover, the other elements (Sn, Zr, and Hf) have similar tendencies as Ti for SRO in the high temperature region, emphasizing that the SRO is not purely driven by Ti and Zr. This further emphasizes that SRO is strongly inhibited in the four-component material relative to the two-component material. 

\begin{figure*}[t]
  \centering
  \includegraphics[width=0.9\textwidth]{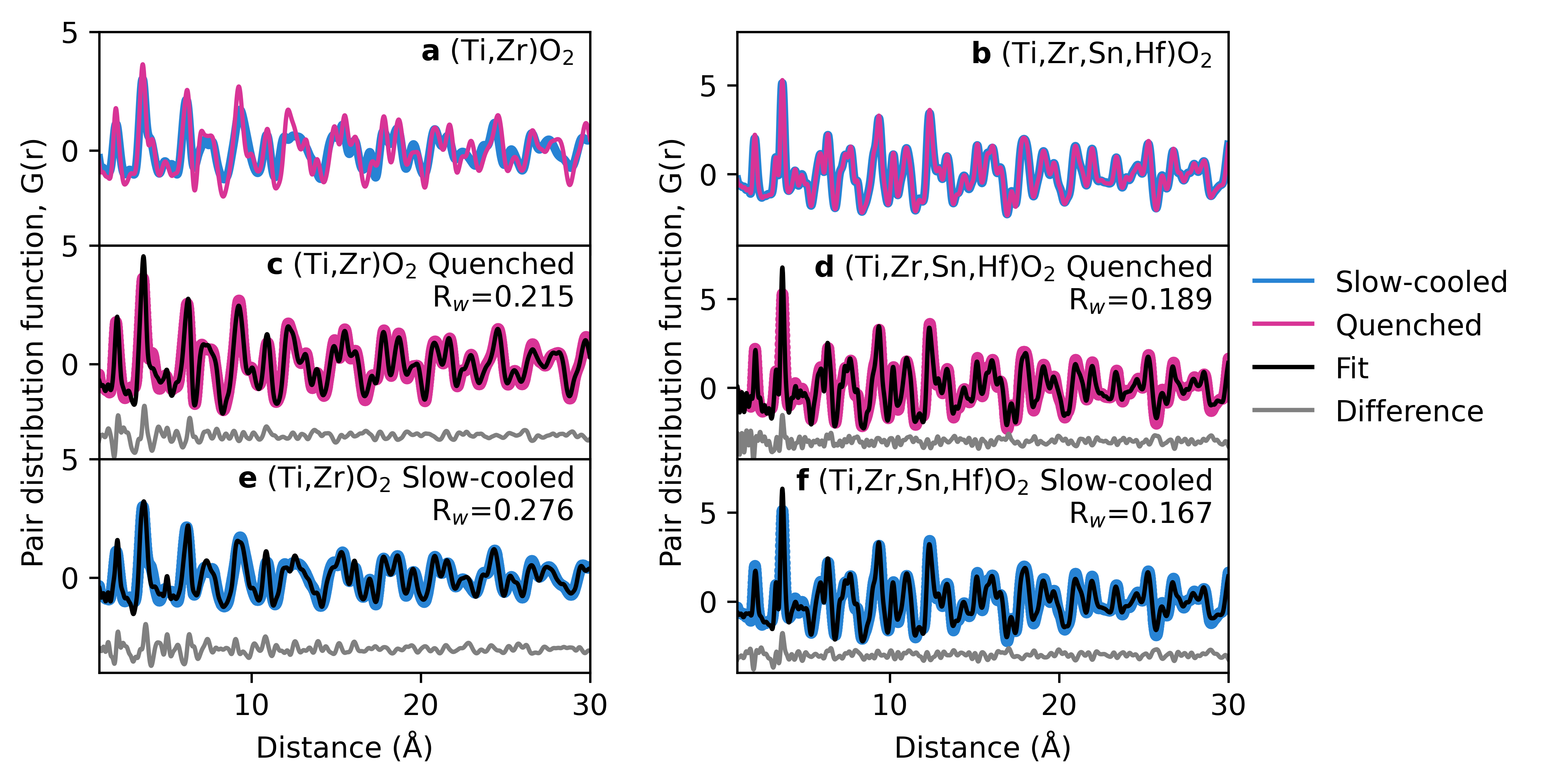}
  \caption{\textbf{PDF small-box fitting of total x-ray scattering.} Direct comparison of PDFs for the quenched and slow-cooled version of the (a) two-component compound and (b) four-component compound. Pair distribution functions, small-box fits, and differences for the (c) quenched (Ti,Zr)O$_2$, (d) quenched (Ti,Zr,Hf,Sn)O$_2$, (e) slow-cooled (Ti,Zr)O$_2$, and (f) slow-cooled (Ti,Zr,Hf,Sn)O$_2$. The fits are better for the more uniform four-component compounds, and also better for the quenched (Ti,Zr)O$_2$ sample than the slow-cooled (Ti,Zr)O$_2$ sample due to less short range ordering. The mismatch is stronger at low r-values, and the region of severe mismatch extends to higher r-values for the slow-cooled (Ti,Zr)O$_2$ sample which has the most short-range ordering.}
  \label{fgr:PDFfit}
\end{figure*}

\subsection{Detection of SRO with total scattering}

Polycrystalline samples of (Ti,Zr)O$_2$ and (Ti,Zr,Hf,Sn)O$_2$ were synthesized using a conventional solid-state synthesis method. Portions of each sample were then given two distinct heat treatments: rapid cooling through quenching in air or slow-cooling, as described in the methods. These differing heat treatments were performed to produce samples with different levels of SRO and local distortions. In particular, based on thermodynamic considerations as well as the Monte Carlo simulations discussed above, we can assume that the quenched sample will achieve a more random cation distribution while the slowly-cooled sample will remain in thermal equilibrium for longer and therefore is more likely to form energetically preferred local orders. Powder x-ray diffraction measurements do not reveal any global changes in symmetry as a result of these different cooling protocols, but subtle variations in lattice parameters are observed as presented in Table \ref{tab:PDFfit}. In the case of the two-component (Ti,Zr)O$_2$, these variations have previously been demonstrated to originate from the SRO of Ti and Zr, which is particularly associated with a contraction of the $b$ lattice parameter \cite{Wang1997}.

In order to directly probe the effect of these different thermal treatments on the level of SRO, we performed total scattering measurements on each sample. The room temperature scattering data is transformed to pair distribution functions (PDFs) $G(r)$, which are displayed in Figure \ref{fgr:PDFfit}(a) for the two-component sample and (b) for the four-component sample. It is immediately apparent that the heat treatment has a significant effect on the two-component sample, but practically no effect in the case of the four-component sample, where the two patterns are overlapping almost identically. The specific differences observed for the two-component sample involve changes in width, positions, and intensity of the peaks associated with certain features in $G(r)$. Comparing the two-component and four-component samples, one can also see that the overall magnitude of the PDF is increased in the latter, which can be accounted for by considering its fraction of heavier elements, which have a larger x-ray scattering cross-section.

The PDFs were fitted using a small-box model, with results shown in Figure \ref{fgr:PDFfit}(c-f) with the complete refined structural parameters shown in the supporting information. These fits assume an undistorted $\alpha$-PbO$_2$ structure where the cation form factor is a stoichiometric mixture of Ti and Zr in the case of the two-component material and Ti, Zr, Hf, and Sn in the case of the four-component sample. While good qualitative agreement is observed for all data sets, some quantitative differences are found. As indicated by the goodness-of-fit metric $R_w$, fit quality is uniformly better for the four-component sample than the two-component one. Zeroing in on the two-component sample, we also see that the fit quality is better for the quenched sample (Fig.~\ref{fgr:PDFfit}(c)) than the slow-cooled (Fig.~\ref{fgr:PDFfit}(e)). This trend aligns precisely with the expected level of SRO present in the sample, with the sample with the highest level of SRO (slow-cooled (Ti,Zr)O$_2$)) having the greatest local structural deviations from the average $\alpha$-PbO$_2$ structure and the correspondingly largest (worst) $R_w$. Close inspection of the (Ti,Zr)O$_2$ fits reveal specific areas of mismatch with the calculated PDF, particularly at the smallest distances where SRO is expected to be most influential. The fit quality in the case of the four-component sample does not significantly differ between the quenched and slow-cooled treatments (Fig.~\ref{fgr:PDFfit}(d,f)), which we ascribe to its lower propensity to exhibit SRO.

Inspection of the refined lattice parameters, which are presented in Table \ref{tab:PDFfit}, gives further indirect evidence on the presence of SRO in these samples. Looking first at (Ti,Zr)O$_2$, we see that the $b$ lattice parameter contracts significantly, by 2\%, as the result of slow-cooling. Previous experimental work has demonstrated that the short-range order induced in (Ti,Zr)O$_2$ by slow-cooling consists of ordered slabs of TiO$_6$ and ZrO$_6$ octahedra stacked along the $a$ axis leading to the contraction of the $b$ axis \cite{Wang1997}. This SRO has a less significant effect on the $a$ and $c$ lattice parameters. Meanwhile, in the case of (Ti,Zr,Hf,Sn)O$_4$, we find that the lattice parameters in the quenched and slow-cooled conditions are indistinguishable within the error of our analysis. We, therefore, conclude that the four-component sample is largely free from SRO due to its ordering transition occurring at experimentally inaccessible temperatures. 

\begin{table}[t]
\centering
\caption{Lattice parameters obtained from the fitting of pair distribution function data. Slow-cooling induced SRO leads to a significant contraction of the $b$ lattice parameter in the case of the two-component sample but not in the four-component sample. The bracketed number represents the standard deviation in the last digit.}
\noindent
\begin{tabular}{lcccc}
\toprule
&\multicolumn{2}{c}{(Ti,Zr)O$_2$}&\multicolumn{2}{c}{(Ti,Zr,Hf,Sn)O$_2$}\\
	&	 Quenched		&	Slow-cooled		&	 Quenched		& Slow-cooled		\\
 \hline
$a$ (\AA)	&	4.820(4)	&	4.826(5)	&	4.836(2)	&	4.836(3)	\\
$b$ (\AA)	&	5.493(5)	&	5.402(6)	&	5.639(3)	&	5.633(3)	\\
$c$ (\AA)	&	5.054(4)	&	5.062(7)	&	5.138(2)	&	5.136(2)	\\
\toprule
\end{tabular}
\label{tab:PDFfit}
\end{table}

\section{Summary and Outlook}
In this work, we have compared SRO and local distortions between a two-component and a four-component entropy stabilized oxide. The experimentally observed ordered phase of the two-component model system is well described by our computational approach,  giving us confidence in the reliability of our findings for the four-component system, for which comparable experimental data is not available. The local distortions are similar in the two materials, governed by ideal cation-oxygen bond lengths. In terms of SRO, the calculations show that the order-disorder transition temperature for the four-component system is lowered to a temperature where cation mobility is low. For this reason, the ordering achieved in the Monte Carlo simulation in the four-component sample is not experimentally observed. On the other hand, the calculations also show that SRO might be present in the two-component compound even at synthesis temperatures. The pair distribution function analysis corroborates this view. We conclude that SRO effects are expected to be less prevalent as more components are added in, provided the pairwise interaction strengths remain comparable between the constituents, while local distortions are less affected by the number of components.

One question raised by this work is how generalizable these trends are to other families of high entropy materials. Our finding that local distortions are similar around the same cation independent of the number of constituents is in agreement with other studies on high entropy alloys, and can be expected to be true for most multicomponent materials~\cite{Owen2018}. This can be understood from the fact that a local reorganization around the cation lowers the energy significantly and the energy barrier to do so it small or non-existent such that kinetic trapping effects are less relevant. Such distortions do not affect neighbors to a large extent, particularly in the presence of the oxygen sublattice. This behavior could be different in materials with more collective interactions, such as Jahn-Teller distortions in ferroelectrics.

The suppression of SRO when increasing the number of elements while keeping the energetics similar is also observed in high entropy battery materials \cite{Squires2023, Lun2021}, where it is found to be advantageous. This suppression can be understood from simple considerations of the Gibbs free energy: if the enthalpic contributions $\Delta H$ stay the same but the entropy $\Delta S_{config}$ is increased then the transition temperature must necessarily decrease proportionally. If the additional components have larger enthalpic contributions, this argument no longer holds. The more elements are in a multicomponent material, the higher the probability that there will be some enthalpic contribution that is stronger than the others \cite{Senkov2015}. The fact that the enthalpic contributions of the two-component and four-compounent oxides discussed here are similar may be the exception rather than the norm. Related to the propensity for SRO, it is also important to emphasize that the materials studied here have an orthorhombic crystal structure, which necessitates that the interaction parameters are directional. Most high entropy oxides meanwhile tend to crystallize in high symmetry space groups~\cite{Aamlid2023Understanding}. The potential for anisotropic interactions makes non-cubic multicomponent materials more susceptible to ordering phenomena.

\section*{Acknowledgements}
This work was supported by the Natural Sciences and Engineering Research Council of Canada, the CIFAR Azrieli Global Scholars program, and the Alfred P. Sloan Foundation. This research was undertaken thanks in part to funding from the Canada First Research Excellence Fund, Quantum Materials and Future Technologies Program.  Part of the research described in this paper was performed at the Canadian Light Source, a national research facility of the University of Saskatchewan, which is supported by the Canada Foundation for Innovation (CFI), the Natural Sciences and Engineering Research Council (NSERC), the Canadian Institutes of Health Research (CIHR), the Government of Saskatchewan, and the University of Saskatchewan. Mario U. Gonz\'alez-Rivas and Solveig S. Aamlid acknowledge the receipt of support from the CLSI Student Travel Support Program.\\

\section*{Supporting information}
The supporting information contains the full cluster-expansion for (Ti,Zr)O$_2$ (Figure S1), the effective cluster interaction parameters (Tables S1-S3), and the structural parameters from the pair distribution functions (Table S4).

\section*{Data availability}
The data are available from the corresponding author upon reasonable request.

\section*{Author contributions}
Sam Mugiraneza and Solveig S. Aamlid prepared the samples. G.~King, Mario U. Gonz\'alez-Rivas and Solveig S. Aamlid performed the PDF measurements at the Canadian Light source. Calculations, data analysis, and visualization were performed by Solveig S. Aamlid with input from J\"org Rottler and Alannah M. Hallas. The manuscript was written by Solveig S. Aamlid, J\"org Rottler, and Alannah M. Hallas with input from all the authors.

\section*{Competing interests}
The authors declare no competing interests.

\begin{footnotesize}
\section{Methods}

\noindent \textbf{Density functional theory calculations:} The Vienna Ab initio Simulation package~\cite{Kresse1993, Kresse1994, Kresse1996Efficiency, Kresse1996Efficient} (VASP) version 5.4.4 with the projector augmented wave~\cite{Blochl1994, Kresse1999} (PAW) method was used for the Density Functional Theory (DFT) calculations. The VASP supplied PBE~\cite{PBE1996} PAW potentials version 5.4 were used, including 12 electrons for Ti, Zr, and Hf, 14 electrons for Sn, and 6 electrons for oxygen valence shells. The PBEsol+U~\cite{PBEsol2008} exchange correlation functional was used, with Hubbard U values of 4.35, 3.35, and 2.7 for the d-electrons on Ti, Zr, and Hf respectively; 0 eV for Sn and oxygen. We used a 2x2x2 supercell of the ground state crystal structure of each binary oxide to calculate these U-values, following a linear response approach ~\cite{Cococcioni2005}. In this approach, the charge density is first calculated with a U-value of 0 for all cations, next a U-value is imposed for one single cation and a non-selfconsistent (NSC) calculation using the original charge density is performed, finally a self-consistent (SC) calculation using the same U-value but allowing the charge density to change is performed. The SC and NSC calculations are repeated for a few U-values around 0, and the number of d-electrons on the perturbed site is recorded. The number of d-electrons as a function of the applied U value is linearly interpolated to find the SC and NSC response functions respectively. The U value is found as the difference in the inverse slope between the SC and NSC functions. The whole procedure was then repeated starting from a new charge density where all the cations have the same U-value until the calculated U-value changes less than 0.1 eV between each repetition. An energy cutoff of 700 eV was used, together with an electronic convergence threshold of 10$^{-8}$ eV and a threshold of maximum force of any one ion of 10$^{-4}$ eV/{\AA}. Gaussian smearing with a width of 0.01 eV and $\Gamma$-centered k-meshes with a spacing of 0.4 {\AA}$^{-1}$ were used. The $\Gamma$-only version of VASP was used for the 144 atoms SQS cells. All degrees of freedom including atomic positions, cell size, and cell shape were allowed to change during relaxations. The atomic simulation environment was used to manage the calculations~\cite{ase-paper}.\\

\noindent \textbf{Special quasi-random structures:} SQS cells were generated using the mcsqs code which is part of the Alloy Theoretic Automated Toolkit (ATAT)~\cite{VanDeWalle2013}. 144 atom SQS supercells were generated. 2-body clusters up to 6 \AA \, and 3-body clusters up to 4 \AA \, were used. SQS cells with a perfect match (all correlations are 0) were found for the two-component compound. No such cell was found for the more computationally demanding four-component compound after running 28 instances for 24 hours, however most pairs have correlation values smaller than 0.1. A typical energy difference between different disorder realizations in the four-component system is 2 meV per atom, and the mean and variance in bond length distributions is comparable between the different disorder realizations. In the main body of this work, an average over 28 SQS cells with dimensions 3x2x2 is used for the four-component system, while an average over one 3x2x2, one 2x3x2 and one 2x2x3 cell were used for the two-component system. The continuous symmetry measure (CSM) values were found for one of the two-component and one of the four-component SQS cells using the ChemEnv tool using standard setting and ensuring that all the cations are found in an octahedral environment \cite{Waroquiers2020}.\\

\noindent \textbf{Effective cluster interactions:} 
Pitike et al. \cite{Pitike2020} showed that the first nearest cation neighbor pairwise interaction parameters were sufficient to replicate the demixing behavior of Cu and Zn in the prototype rock-salt (Mg,Co,Ni,Cu,Zn)O. 
Due to the orthorhombic nature of the $\alpha$-PbO$_2$ structure, there are three different types of nearest cation neighbor interactions in these compounds. In order to assign an energy to each pairwise interaction, three unit cells with ordering along three different directions, in addition to the pure compounds to establish the zero-point energies, were calculated (5 in total for the two-component and 22 for the four-component). The CLEASE software with the binary linear implementation of cluster expansions was used to keep track of the calculations and clusters and solving linear equations to find the effective cluster interactions (ECIs) \cite{Chang2019, Zhang2016}. This approach is effectively a severely truncated cluster expansion. A conventional cluster expansion was performed for the two-component system, as outlined in the supporting information. \\

\noindent \textbf{Monte Carlo simulations:}
Metropolis Monte Carlo (MMC) simulations are also implemented in CLEASE, where the lattice and corresponding ECIs are the inputs. A 6x6x6 unit cell (2592 atoms, 864 cations, one sweep includes 864 proposed moves) was initialized. A simulated annealing protocol in 5 degree intervals from 2000 K to 5 K was performed, with 100 sweeps for equilibration and 1000 sweeps for sampling at each step. 
The energy was recorded on every sweep, and the acceptance rate, cluster functions, average energy, heat capacity, and energy variance are recorded at each temperature. \\

\noindent \textbf{Thermodynamic integration:}
The Gibbs free energy can be found from the internal energy from the Monte Carlo simulations provided that a reference state where G is known exists. In this work, we make the assumption that at the lowest temperature $T_0$ in the MMC simulations there is a perfectly ordered state with zero entropy such that $G(T_0) = U(T_0)$, which is reasonable but not necessarily always true if the MMC simulations never reach a perfectly ordered state. (or as another option, a completely disordered high-temperature state). The internal energy of the disordered systems in this work is simply the energy from the MMC simulation.

The equation to be solved is \cite{Chang2019}

\begin{equation}
\beta G = (\beta G)_{ref} + \int_{\beta_{ref}}^{\beta} d\beta ' U (\beta '),
\label{eqn:thermondynamic integration}
\end{equation}

\noindent where $\beta = (k_B T)^{-1}$. Practically, this is solved numerically in a stepwise fashion starting from the lowest equilibrated temperature and evaluated at each step

\begin{multline}
    G(\beta_i)=\beta_{i-1}G(\beta_{i-1})+\frac{(U(\beta_{i-1})+U(\beta_i))(\beta_i-\beta_{i-1})}{2\beta_i}
\label{eqn:solve}
\end{multline}

\noindent The relationship between Gibbs free energy, the internal energy, and configurational entropy $\Delta S_{config}$ is

\begin{equation}
\Delta S_{config}(T)=\frac{U(T)-G(T)}{T},   
\end{equation}

\noindent which is what allows for the direct calculation of the non-ideal configurational entropy from the Monte Carlo simulations.\\

\noindent \textbf{Solid state synthesis and thermal treatments:}
All samples were prepared by solid-state synthesis. Elemental oxide precursors were mixed in an agate mortar, uniaxially pressed to pellets, and sintered in air at 1400-1450 \degree C twice for 24 hours each time with an intermediate regrinding \cite{Aamlid2023Phase}. Two types of heat treatments were used to give the samples different degree of ordering: quenching in air by taking the sample out of the furnace at the sintering temperature or a slow-cooling procedure where the samples are cooled at 5 \degree C min$^{-1}$ from 1400 to 1250 \degree C, 0.1 \degree C min$^{-1}$ from 1250 to 1000 \degree C, and 1 \degree C min$^{-1}$ from 1000 \degree C to room temperature~\cite{Kim2001}.\\ 


\noindent \textbf{Total scattering:} 
The total scattering data was collected at the High Energy Wiggler Beamline in the Brockhouse diffraction sector at the Canadian Light Source. The samples were loaded in 0.8 mm diameter kapton capillaries and loaded  in transmission geometry. The beam energy was 65.02 keV and the distance to the Perkin Elmer area detector was about 16 cm. A Nickel standard was used to calibrate the detector distance, integration, and the Q-dependent broadening and dampening in the PDF analysis. The collected 2D images were integrated and transformed to pair distribution function data with a Q$_{max}$ of 24~\AA$^{-1}$ using GSAS-II \cite{Toby2013GSASII}. The analysis was done using PDFgui \cite{Farrow2007PDFgui}. For the small-box modelling, a \textit{Pbcn} space group was used and lattice constants, symmetry-allowed atomic positions, and anisotropic thermal parameters were refined for the cations and isotropic thermal parameters for the oxygen. 

\end{footnotesize}

\bibliography{bibliography}

\end{document}